\documentclass[journal,12pt,onecolumn, draftclsnofoot]{IEEEtran}
\usepackage[dvips]{graphicx}
\usepackage{multirow}
\usepackage{caption}
\usepackage{multicol,lipsum}
\usepackage{mathtools, cuted}
\usepackage{comment}
\usepackage{amsmath}
\usepackage{amsthm}
\usepackage[flushleft]{threeparttable}
\usepackage{array}
\usepackage{booktabs}
\usepackage{dirtytalk}
\newcommand{\PreserveBackslash}[1]{\let\temp=\\#1\let\\=\temp}
\newcolumntype{C}[1]{>{\PreserveBackslash\centering}p{#1}}
\newcolumntype{R}[1]{>{\PreserveBackslash\raggedleft}p{#1}}
\newcolumntype{L}[1]{>{\PreserveBackslash\raggedright}p{#1}}

\usepackage{bm}
\usepackage{graphicx,subfigure}
\usepackage{algorithmic, algorithm}
\usepackage{cite}
\usepackage{amsmath}
\usepackage{amssymb}
\usepackage{psfrag}
\usepackage{empheq}
\usepackage{latexsym, amsmath, subfigure, color, amsfonts, amssymb,graphicx}
\usepackage{wrapfig} 

\IEEEoverridecommandlockouts
\usepackage{tabularx}
\makeatletter
\def\hlinewd#1{%
\noalign{\ifnum0=`}\fi\hrule \@height #1 %
\futurelet\reserved@a\@xhline}
\makeatother
\begin{document}
\title{Deep Learning for Massive MIMO Channel State Acquisition and Feedback}
\author{Mahdi Boloursaz Mashhadi, and Deniz G\"{u}nd\"{u}z\\
\IEEEauthorblockA{Dept. of Electrical and Electronic Eng., Imperial College London, UK\\
Email: \{m.boloursaz-mashhadi, d.gunduz\}@imperial.ac.uk}}

\maketitle
\begin{abstract}
Massive multiple-input multiple-output (MIMO) systems are a main enabler of the excessive throughput requirements in 5G and future generation wireless networks as they can serve many users simultaneously with high spectral and energy efficiency. To achieve this massive MIMO systems require accurate and timely channel state information (CSI), which is acquired by a training process that involves pilot transmission, CSI estimation, and feedback. This training process incurs a training overhead, which scales with the number of antennas, users, and subcarriers. Reducing the training overhead in massive MIMO systems has been a major topic of research since the emergence of the concept. Recently, deep learning (DL)-based approaches have been proposed and shown to provide significant reduction in the CSI acquisition and feedback overhead in massive MIMO systems compared to traditional techniques. In this paper, we present an overview of the state-of-the-art DL architectures and algorithms used for CSI acquisition and feedback, and provide further research directions.

\end{abstract}
\section{Introduction}\label{intro}
Massive multiple-input multiple-output (MIMO) systems are an important component of 5G and future generation wireless networks due to their ability to serve many users simultaneously with high spectral and energy efficiency. The main idea in massive MIMO is to equip base stations (BSs) in wireless networks with large arrays of cooperating antennas to facilitate spatial multiplexing of many user equipments (UEs) within the same time-frequency resources. Since the number of antennas at the BS is typically assumed to be significantly more than the number of users, a large number of degrees of freedom are available in the downlink, which can be used to shape the transmitted signals in a specific direction or to null interference. This yields a beamforming gain that translates into increased energy efficiency, reduced interference, or improved coverage. In the uplink, each single-antenna user in a massive MIMO system can scale down its transmit power proportionally to the number of antennas at the BS while maintaining the same performance as the corresponding single-input single-output (SISO) system. This leads to higher energy efficiency, which is a major benefit in next generation wireless networks, where excessive energy consumption is a growing concern. On the other hand, if adequate transmit power is available, a massive MIMO system can significantly expand its coverage compared to a single-antenna system.

In communication systems, channel state information (CSI) is required at the receiver to be able to decode the information transmitted over a time-varying channel. CSI is acquired by a training process which involves pilot transmission and CSI estimation at the receiver. This imposes a training overhead on the communication system which scales up with the number of antennas, receivers and subcarriers. In massive MIMO systems, to achieve the aforementioned performance gains, accurate and timely CSI is required both at the BS and the UEs. Availability of downlink CSI at a massive MIMO BS is crucial to enable beamforming and achieve spatial multiplexing gains. Reducing the training overhead in massive MIMO has been a major topic of research since the emergence of the concept. 

\begin{figure*}
\centering
\includegraphics[scale=0.45]{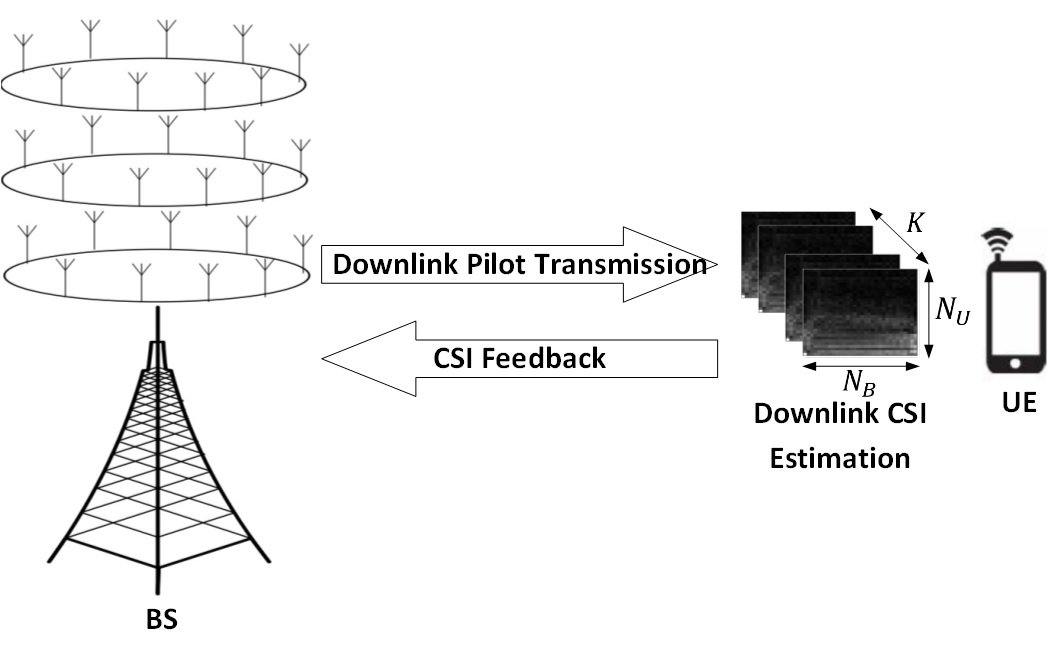}
\caption{Massive MIMO downlink training process in FDD mode.}
\label{fig1}       
\end{figure*}

Massive MIMO was originally introduced in a time division duplex (TDD) setting where the uplink and downlink channels are separated in time \cite{MMIMO1, MMIMO2}. In the TDD mode of operation, due to uplink/downlink channel reciprocity, which holds under certain conditions \cite{MarzettaCSI}, downlink CSI does not induce extra training overhead. However, motivated by spectrum regulation issues, FDD operation gained significant interest \cite{FDDMMIMO1, FDDMMIMO2}, and there has been a long-standing debate on the relative performance of  TDD and FDD schemes \cite{TDDvsFDD1, TDDvsFDD2, FDDCSI3}. Although the FDD  scheme is favourable due to its improved coverage and reduced interference, these benefits come at the price of increased complexity of the training process for FDD massive MIMO. Unlike in TDD, in the FDD mode of operation, the uplink and downlink channels are separated in frequency, and hence, they are not reciprocal. Consequently, in FDD massive MIMO, downlink CSI need to be first estimated at each UE, and then fed back to the BS through the uplink channel, which significantly increases the CSI overhead. Fig. \ref{fig1} depicts the downlink training process in FDD mode.

For smaller number of BS antennas, simple vector quantization (VQ) along with exhaustive search may work sufficiently well for MIMO CSI compression and feedback. In the fourth generation long term evolution (4G-LTE) advanced standard, a 4-bit channel quality index (CQI) and the pre-coding matrix indicator (PMI) are fed back to the BS to reveal the CSI \cite{CQI}. However, with the increased number of massive MIMO antennas, CSI dimensions increase drastically and the traditional VQ-based approaches are no longer practical. This has encouraged great interest in more efficient training and compression techniques. Initial efforts in this direction followed a model-based approach assuming sparse or low-rank models on the CSI matrix. However, a sparse model on the channel is less accurate when MIMO dimensions are not sufficiently large, which degrades the performance of sparsity-based techniques. The same discussion holds for low-rank based techniques \cite{LRest1, LRest2}, where there is a model mismatch. These approaches do not take into account the inherent statistical correlations and structures beyond sparse or low-rank patterns. Moreover, sparse and low-rank reconstruction techniques are computationally demanding iterative algorithms, which may further limit their practical implementation.


NN-based approaches have recently shown significant improvements over their model-based counterparts in various wireless communication problems \cite{MLintheAir, Hoydis, Hoydis2, deepJSCC}. Data-driven approaches have also been proposed for massive MIMO channel estimation and feedback, in order to exploit the common structures observed in typical massive MIMO CSI matrices. Data-driven approaches train neural network (NN) structures over large datasets of CSI matrices to capture these structures and use them to reduce the CSI acquisition overhead. 

Consider the following massive MIMO channel matrix $\mathbf{H}(\tau) \in \mathbb{C}^{N_U \times N_B}$ in the delay domain: 
\begin{align}\label{Hform1}
    \mathbf{H}(\tau)=\sqrt{\frac{N_U N_B}{L}}\sum^{L}_{l=1} \alpha_l \delta(\tau-\tau_l) \mathbf{a}_U(\theta_l) \mathbf{a}_B^H(\phi_l),
\end{align}
where $N_B$ and $N_U$ denote the number of antennas at the BS and UE, respectively, $L$ is the number of multi-path components with $\alpha_l$ denoting the propagation gain of the $l$th path. Also, $\mathbf{a}_B$ and $\mathbf{a}_U$ are the array response vectors for the BS and user with  $\theta_l$ and $\phi_l$ denoting the azimuth angles of arrival and/or departure (AoA/AoD), respectively, and $(\cdot)^H$ denotes the conjugate transpose operation. For uniform linear arrays, we have 
\begin{align}\label{Hform2}
    \mathbf{a}_U(\theta_l)&=[1, e^{-j\frac{2\pi d}{\lambda} \sin{\theta_l}}, \cdots, e^{-j\frac{2\pi d}{\lambda} (N_U -1) \sin{\theta_l}}]^T / \sqrt{N_U},\\
    \mathbf{a}_B(\phi_l)&=[1, e^{-j\frac{2\pi d}{\lambda} \sin{\phi_l}}, \cdots, e^{-j\frac{2\pi d}{\lambda} (N_B -1) \sin{\phi_l}}]^T / \sqrt{N_B},
\end{align}
where $d$ and $\lambda$ denote the distance between adjacent antennas and the carrier wavelength, respectively. Equivalently, the MIMO channel matrix at the $k$th subcarrier in OFDM, $\mathbf{H}_k \in \mathbb{C}^{N_U \times N_B}$, is given by

\begin{align}\label{Hform3}
    \mathbf{H}_k=\sqrt{\frac{N_U N_B}{L}}\sum^{L}_{l=1} \alpha_l e^{-j2\pi \tau_l f_s \frac{k}{K}} \mathbf{a}_U(\theta_l) \mathbf{a}_B^H(\phi_l),
\end{align}
where $f_s$ denotes the sample rate and $K$ is the total number of subcarriers. 

According to (\ref{Hform3}), the CSI values for nearby users, sub-carriers and antennas are correlated due to similar propagation paths, gains, delays and AoDs/AoAs. Apart from the correlations governed by (\ref{Hform3}), there exists inherent characteristics in MIMO environments due to specific user distributions, scattering parameters, geometry, materials, etc., that cause common structures among MIMO CSI matrices. We note that the joint statistics of the channel gains across antennas, subcarriers and users is extremely complex. Even if accurate models are known on the statistics in (\ref{Hform3}), identifying a lossy compression scheme to optimally exploit structures and correlations in (\ref{Hform3}) is challenging. On the other hand, NNs are extremely powerful in learning complex distributions and exploiting them for various classification/regression (supervised learning) or compression (unsupervised learning) tasks. NNs can be used to learn the common structures and inherent correlations to leverage them for efficient CSI estimation, compression and feedback, reducing the overall MIMO training overhead. 

Success of data-driven approaches depends critically on the datasets used to train the NN models. Unlike some more popular applications of NNs, rich and standardized datasets of CSI measurements in actual massive MIMO scenarios do not yet exist. However, there exists MIMO channel models that have proved to be very accurate in statistically modeling actual CSI measurements in practical MIMO scenarios. Among these are the third generation partnership project (3GPP) spatial channel model (SCM) \cite{3GPP1}, WINNER II \cite{WINNER} and COST 2100 \cite{COST2100}. Unfortunately, existing results in the literature use different channel models to generate CSI datasets ranging from the simple formula in (\ref{Hform3}) to more sophisticated channel models like COST 2100, 3GPP TR 38.901 release 15 \cite{3GPP} or the DeepMIMO ray-tracing propagation model \cite{DeepMIMO}. The most widely used channel model so far has been COST2100, which will also be used in this paper. We would like to emphasize that different datasets hamper the comparison of different results and there is a pressing need for standard datasets.

This paper provides an overview of how NNs can be used in massive MIMO systems to improve the performance of CSI acquisition and feedback while reducing both the complexity and overhead. In the following sections, we shall review recently proposed data-driven approaches for CSI estimation, compression and feedback, and provide suggestions for future research.

\section{MIMO Channel Estimation by DL}

\label{sec1}
Consider uplink MIMO training where the user transmits a block of $P$ pilot signals, denoted by $\mathbf{X} \in \mathbb{C}^{N_U \times P}$, which is known at both the UE and the BS. The BS needs to estimate the channel matrix $\mathbf{H} \in \mathbb{C}^{N_B \times N_U}$ from received measurements $\mathbf{Y} \in \mathbb{C}^{N_B \times P}$, given by
\begin{align}\label{Chest1}
    \mathbf{Y}= \mathbf{H} \mathbf{X} + \mathbf{Z},
\end{align}
where $\mathbf{Z} \in \mathbb{C}^{N_B \times P}$ is the complex additive white Gaussian noise (AWGN). 

Standard channel estimation techniques are typically based on linear minimum mean square error (LMMSE) estimation method. A common assumption in LMMSE-based channel estimation techniques is that the pilot length is larger than the number of transmit antennas, which may be prohibitive in downlink training of massive MIMO systems ($P \ge N_B$). For downlink massive MIMO channel estimation, where $N_B$ is large, it is challenging to ensure $P \ge N_B$ not only because it shall increase the training overhead and computational complexity for channel estimation, but also because a large $P$ may even exceed the channel coherence interval. If this assumption does not hold, LMMSE-based channel estimation performance degrades significantly.

Many previous works take a model-based estimation approach assuming sparse \cite{CSest1, CSest2, CSest3} or low-rank \cite{LRest1, LRest2} models on the channel matrix. Sparsity of the channel in the angular-delay domain has been assumed in \cite{CSest1, CSest2, CSest3}, where compressive sensing based reconstruction techniques are used to reduce the pilot length and training overhead. Sparsity based techniques can decrease the pilot length required to sense and estimate the channel by an order of magnitude \cite{CSest4} compared to a simple exhaustive search approach. However, as mentioned earlier, these techniques rely on the sparse or low-rank properties of the channels, which may not be very accurate and do not take into account the inherent statistical correlations and structures beyond sparse or low-rank patterns.

This motivates the use of data-driven approaches based on NNs to learn these complex structures and correlations. The authors in \cite{DLest1, shlezinger2019deep} use convolutional NNs to improve the quality of a coarse initial estimate of the channel matrix by exploiting temporal and inter-frequency correlations. Let $\mathbf{H}_k(n) \in \mathbb{C}^{N_B \times N_U}$ denote the MIMO channel matrix for the $k$th subcarrier at temporal slot $n$, where the channel is assumed constant during each slot, which corresponds to the channel coherence time. A coarse initial estimate of $\mathbf{H}_k(n)$ is given by $\mathbf{R}_k(n)= \mathbf{Y}_k(n) \mathbf{X}_k(n)^{\dagger}$, where $\mathbf{X}_k(n)^{\dagger}$ denotes the pseduo-inverse of the pilot signals transmitted over the $k$th subcarrier at time $n$. The authors form large tensors by concatenating $\mathbf{R}_k(n)$'s along time and frequency dimensions, and then apply multi-dimensional convolution kernels on it. During training, these kernels capture temporal and inter-frequency correlations, and can be exploited to provide accurate estimates of the channel matrix. This idea outperforms non-ideal minimum mean square error (MMSE) (with estimated covariance matrix) estimation and achieves performance very close to the ideal MMSE (with true covariance matrix) that is very difficult to be implemented in practical situations. The NN architecture used in \cite{DLest1} consists of 12 convolutional layers. There is still much work to be done to design NN architectures with reduced complexity and improved performance to guarantee that the channel estimation task can be carried out rapidly within the channel coherence time. 

On the other hand, many massive MIMO structures use low-resolution analog-to-digital converters (ADCs) to reduce the power consumption and hardware complexity at the BS; and hence, only a coarsely quantized version of $\mathbf{Y}$ shall be available for channel estimation at the BS. For the quantized case, we have
\begin{align}\label{1bit1}
    \mathbf{Y}_q= \mathcal{Q}(\mathbf{H} \mathbf{X} + \mathbf{Z}),
\end{align}
where $\mathcal{Q}(\cdot)$ denotes quantization performed element-wise on the real and imaginary parts of the received signals independently. Low-resolution ADCs incur nonlinear distortion, which poses significant challenges to channel estimation from highly quantized measurements. Hence, efficient estimation techniques from quantized received signals $\mathbf{Y}$ are needed.

With coarsely quantized measurements, the pilot length required for reliable estimation of the channel; and hence, the training overhead increases significantly. Model-based estimation techniques generally minimize a cost function (e.g., maximum likelihood, square error, etc.) iteratively subject to sparsity \cite{LowResCS1, LowResCS2, LowResCS3} or low-rank \cite{LowResLowRank1} constraints on the channel matrix $\mathbf{H}$. Due to the additional non-linearity introduced by quantization, NN-based techniques can be even more beneficial in channel estimation from low-resolution received signals.  

For the extreme case of 1-bit ADCs, reconstruction is possible only up to a scale factor. The initial results reported in \cite{DL1Bit, balevi2019twostage} show that a simple fully-connected network trained in a supervised setting to estimate the channel directly from sign measurements can reduce the required pilot length roughly by an order of magnitude, while achieving similar reconstruction performance in comparison with previous sparse or low-rank based techniques. In \cite{DLFewBit}, the authors consider a mixed-ADC scenario, where several BS antennas are equipped with high resolution ADCs and others with few-bit ADCs to achieve a trade-off between the performance and power consumption. They input an initial least square (LS) channel estimate to a 5-layer fully-connected NN, and show that the NN can learn to utilize the correlation between antennas to improve the estimation performance for the low-resolution branches. The above works utilize inter-antenna correlations for channel estimation; temporal and spectral correlations can be similarly exploited by convolutional kernels.

While fully-connected NNs have been commonly used in previous works and they have the potential to learn and exploit complex joint distributions across all antennas and subcarriers, they do not easily scale with MIMO dimensions and need separate training for different number of antennas, subcarriers, etc. However, as discussed earlier, correlations in typical MIMO channels exhibit locality among antennas and subcarriers, which encourages utilizing convolutional architectures, which can significantly reduce the complexity in both training and inference. This is especially critical in wireless applications, as it is important to acquire an accurate channel estimate within the channel coherence time. Moreover, convolutional kernels, once trained, work for different input dimensions; that is, we do not need to train and use a different NN when the number of antennas in either side of the channel, or the number of subcarriers allocated for communication change. 

\section{DL-based MIMO CSI Reduction and Feedback}\label{sec2}

Once the channel matrix $\mathbf{H}$ is estimated at the UE, it needs to be transmitted back to the BS through a feedback channel, which incurs further overhead. With massive number of antennas and increased bandwidth and  users, the CSI dimensions, and the resulting overhead, increase significantly, which motivate CSI reduction techniques. Traditional CSI compression techniques include vector quantization (VQ), sparsifying transforms (e.g., discrete cosine transform (DCT), Karhunen-Loeve transform (KLT)), principal component analysis (PCA)-based dimensionality reduction and compressed sensing (CS) to compress the CSI using spatio-temporal MIMO channel correlations.

However, as we have mentioned earlier, lossy compression is a challenging task even when the underlying source distributions is known perfectly. While we have a relatively good understanding of the fundamental rate-distortion performance for independent and identically distributed sources in the asymptotic limit, lossy compression for practical sources, such as image, audio, or video, has been a research challenge for many decades. Recently, dimensionality reducing autoencoders have shown significant success for lossy compression of such sources with a data-driven approach. Similarly for the CSI, dimensionality reducing autoencoders have recently been used to efficiently reduce the massive MIMO CSI overhead. These autoencoder architectures can be trained to learn a lower dimensional representation of the original CSI matrix to be transmitted over the feedback channnel with a reduced overhead. An initial study using this autoencoder approach showed significant improvement in comparison with the best performing sparsity-based techniques \cite{wen2018deep}. The authors in \cite{wen2018deep} proposed CSINet, which has since been adopted as a benchmark architecture for performance comparisons by subsequent works. CSINet includes convolutional layers as well as dense layers and Refine-Net architectures. In \cite{DLCSI2, DLCSI3}, the authors combine CSINet and long short-term memory (LSTM) cells to improve upon the basic CSINet architecture by exploiting the temporal correlations in CSI matrices for consecutive time instances.  The authors in \cite{DLCSI4} use the uplink CSI (which is already available to the BS by uplink training) as a side information to further improve CSI reconstruction performance utilizing the correlations between downlink and uplink channels.

These DL-based CSI reduction techniques mainly train an end-to-end auto-encoder structure, assuming ideal feedback of the reduced CSI. However, the estimated CSI (of the downlink channel) is fed back to the transmitter through the uplink channel, which also suffers from noise, interference, and fading. It becomes crucial to design CSI compression and feedback schemes that not only reduce the CSI overhead efficiently, but are also robust against the feedback channel impairments.

There are two main approaches to cope with the limitations in the CSI feedback channel, i.e., the \textit{digital} and \textit{analog} CSI schemes. Digital schemes, which have traditionally received more attention, are based on the separation approach: CSI is first compressed into as few bits as possible and these bits are reliably fed back to the transmitter using a low-rate channel code, which adds redundancy in a way to cope with the channel noise and error in the feedback link. On the other hand, analog CSI follows a joint source-channel coding approach, and directly maps the downlink CSI to the uplink channel input in an unquantized and uncoded manner. Analog approach simplifies the feedback operation as it does not require explicit quantization, coding, and modulation. If the uplink feedback channel is an additive white Gaussian noise channel, and the downlink CSI is Gaussian and perfectly known at the UE, the analog CSI scheme (that incurs zero delay) is optimal in that it achieves the same minimum mean-squared error distortion for the reconstructed CSI at the BS as a scheme that optimally quantizes and encodes the CSI, while incurring infinite delay. The low-latency of the analog CSI scheme makes it a favourable alternative in rapidly changing MIMO channels where the CSI needs to be estimated and fed back to the BS periodically. We shall overview both analog and digital CSI schemes in the presence of feedback channel impairments in the following subsections.

\subsection{Digital CSI feedback}

The earlier autoencoder-based CSI reduction techniques \cite{wen2018deep, DLCSI2, DLCSI4} overlooked the subsequent feedback of the reduced CSI, and mainly focused on the dimensionality reduction by a direct application of the autoencoder architecture. These works are based on the assumption that reducing the dimension of CSI matrix would result in reduced feedback overhead. This is not neccessarily correct since the reduced representation consists of real numbers, which may still need to be compressed further, and the impact of such compression on the final CSI accuracy is not taken into account. Several subsequent works assume that the reduced CSI is quantized before being digitally fed back to the BS. The authors consider simple uniform quantization in \cite{FEDDEL} and non-uniform $\mu$-law quantization in \cite{CSINETPlus}. Since quantization is a non-differentiable function, the gradient cannot pass through it in the backpropagation step of the learning algorithm. This makes it challenging to train digital CSI feedback schemes in an end-to-end manner and requires further considerations to overcome the gradient backpropagation issue. A widely used solution is to set the quantization gradient to a constant, and train end-to-end for a specific number of quantization bits. The authors in \cite{CSINETPlus} add an offset module to the decoder to compensate for the quantization distortion, where the network is trained in multiple stages: end-to-end training without quantization with a larger learning rate, followed by quantization and optimization of the offset module, and finally the offset and decoder are fine tuned by further training with a small learning rate. 

Although the authors in \cite{FEDDEL, CSINETPlus} consider quantization of the reduced CSI to convert it into bits to be transmitted over the feedback link, the simple scalar quantization approach cannot fully exploit the potential correlations remaining among the components of the reduced CSI. Indeed, the quantizer output does not produce equally probable bits; and hence, additional lossless compression of the bits would further reduce the feedback overhead.

\begin{figure*}
\centering
\includegraphics[scale=0.47]{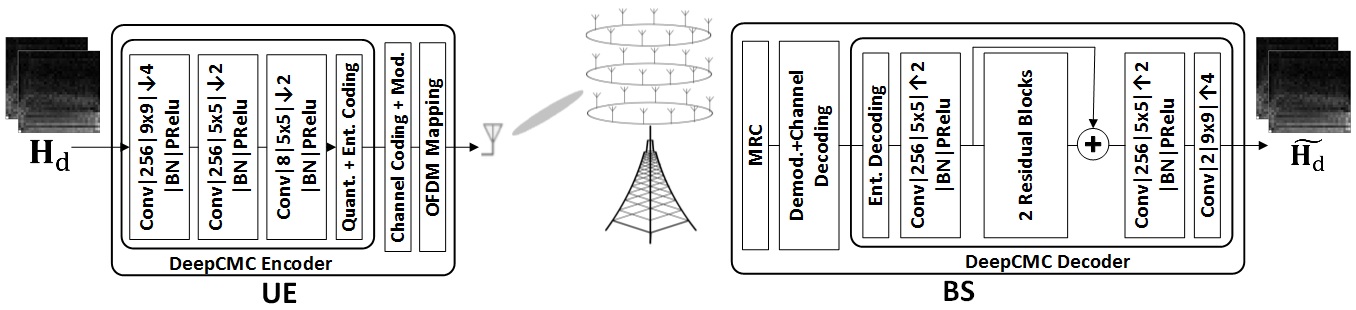}
\caption{DeepCMC for MIMO CSI compression.}
\label{fig2}
\end{figure*}

In \cite{yang2019deep, yang2019deepJ}, a DL-based CSI matrix compression technique, called DeepCMC, is proposed, which employs entropy coding to further compress the quantizer outputs. Fig. \ref{fig2} provides the end-to-end block diagram for a downlink digital CSI feedback scheme based on DeepCMC \cite{yang2019deep, yang2019deepJ}. In this figure, $\mathbf{H}_d$ and $\widetilde{\mathbf{H}_d}$ denote the downlink CSI matrix at the UE and its estimate at the BS, respectively, and the two model input matrices represent $\Re(\mathbf{H}_d)$ and $\Im(\mathbf{H}_d)$. The UE applies a CNN-based feature encoder on $\mathbf{H}_d$ to obtain its low-dimensional representation, which is subsequently quantized and compressed using context-adaptive binary arithmetic coding (CABAC) \cite{marpe2003context}. The resulting bit stream passes through channel coding and digital modulation. The modulation output is then mapped over OFDM subcarriers and transmitted back to the BS over the uplink channel. The BS performs maximum ratio combining (MRC) on the received signals to maximize the SNR and benefit from the diversity in the feedback channel. The resulting signal then passes through the demodulation, channel decoding, entropy decoding, and finally the CNN-based feature decoder to reconstruct $\widetilde{\mathbf{H}_d}$. 

\begin{figure}
\centering
\includegraphics[scale=0.7]{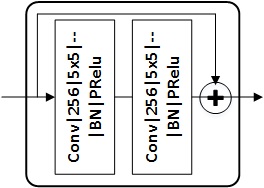}
\caption{The residual block model.}
\label{fig3}
\end{figure}

In the CNN architecture in Fig. \ref{fig2}, \say{Conv$|256|9\times9| \downarrow4|$BN$|$PRelu} represents a convolutional layer with 256 features and kernel size of $9\times9$ followed by downsampling by a factor of 4, batch normalization and parametric rectified linear activation unit (PReLU). As depicted in Fig.~\ref{fig2}, the feature decoder consists of three convolutional layers and two residual blocks with shortcut connections, where \say{+} denotes simple element-wise addition. Fig.~\ref{fig3} illustrates the architecture for each residual block, where \say{$|--$} means the corresponding convolution output is not downsampled. The residual and shortcut structures ease training of the network by  preventing vanishing gradients along the stacked non-linear layers and improve the performance according to our simulation results.

The training cost for DeepCMC is a weighted sum of the mean square error (MSE) of the CSI reconstruction and the quantizer's output entropy. A weight parameter $\lambda$ controls the trade-off between the reconstruction quality and the feedback bit rate, with a larger value resulting in improved MSE at an increased bit rate. For a good quality feedback channel with larger capacity, utilizing a network trained with a larger $\lambda$ results in improved CSI quality at the BS. However, if the feedback channel capacity is smaller than the resulting bit rate, the feedback channel will fail to deliver the CSI. To avoid this, a network trained to work at a lower bit rate (trained with smaller $\lambda$) should be used. Different $\lambda$ values will provide networks that work on different points on a rate-distortion curve. The UE will store different networks, and use the proper one depending on the uplink channel state and the capacity achievable for CSI feedback.

We note that, in contrast to the literature on CSI feedback, which has mainly focused on minimizing the reconstruction error, DeepCMC is trained with a rate-distortion cost that takes into account both the compression rate (in terms of bits per channel dimension) and the reconstruction MSE. As we will see below, this additional compression step leads to a significant improvement in the achieved performance. It also allows adapting the CSI quality to the available feedback channel quality.

Another important benefit of the DeepCMC architecture is that, it is fully convolutional, and has no densely connected layers, which makes it flexible for a wide range of MIMO scenarios with different number of sub-channels and antennas. As shown by the simulation results, although DeepCMC is trained for a specific number of sub-channels and antennas, it generalizes well to other configurations with different number of sub-channels and antennas \cite{yang2019deep, yang2019deepJ}. This is very important for practical implementation of NN-based CSI compression techniques, as otherwise the nodes would have to store a large number of NN parameters for every possible combination of antenna and subcarrier configurations.

\begin{figure}
\centering
\includegraphics[scale=.8]{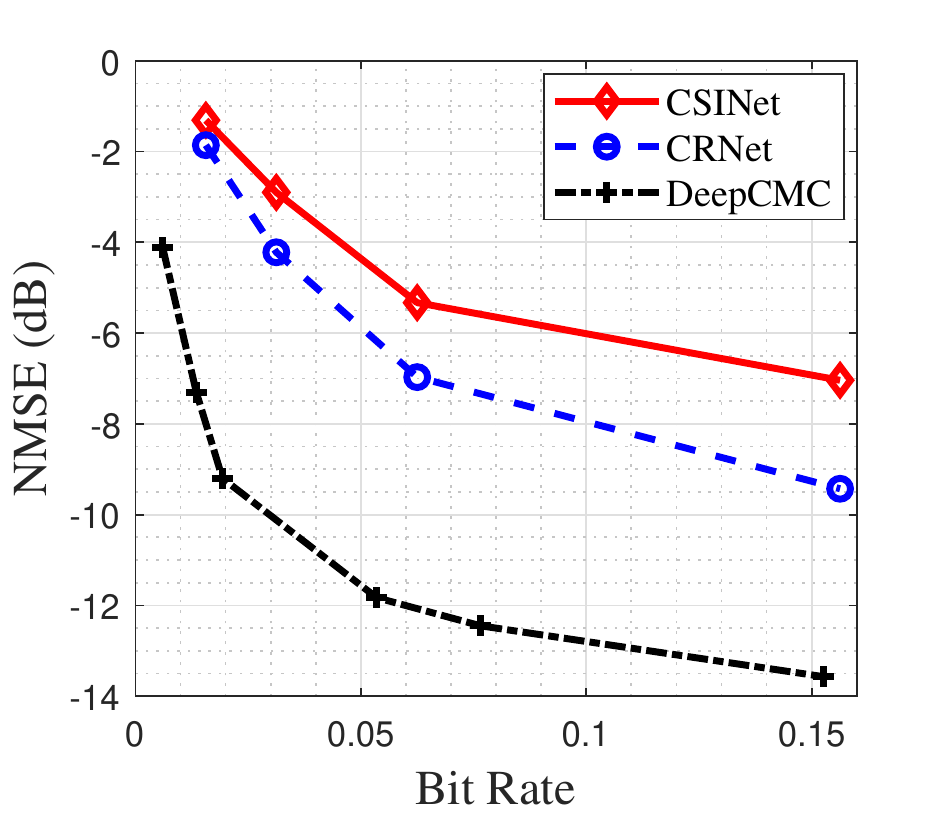}
\caption{Bit rate-NMSE trade-off comparison, $K=256$, $N_B=32$, $N_U=1$.}
\label{curve1}
\end{figure}

In Fig. \ref{curve1}, we present a comparison of the output rate-distortion curves for DeepCMC \cite{yang2019deep, yang2019deepJ}, CSINet \cite{wen2018deep} and CRNet \cite{CRNet}. In this comparison, we use the normalized mean square error defined as $\mathrm{NMSE} \triangleq \mathbb{E}\left[ \frac{ \|\mathbf{H}_d-\widetilde{\mathbf{H}_d}\|_2^2}{ \|\mathbf{H}_d\|_2^2} \right]$. We plot the achieved NMSE, in dBs, as a function of the average number of bits used to encode each CSI entry. Note that the outputs for CSINet and CRNet are feature vectors of type \say{float32}, and hence 32-bit quantization is considered to calculate the resulting bit rate for them. 

For the comparison in Fig. \ref{curve1}, we consider downlink training for a single-antenna user in an FDD MIMO setting. We set $K=256, N_B=32, N_U=1$, and use the COST2100 channel model \cite{COST2100} to generate sample channel matrices for training and testing. We consider the indoor picocellular scenario at 5.3 GHz, where the BS is equipped with a ULA of dipole antennas positioned at the center of a $20\mathrm{m} \times 20\mathrm{m}$ square. The user is placed within this square uniformly at random. All other parameters follow the default settings in \cite{COST2100}. The number of training and testing samples are 80000 and 20000, respectively, and the batch size is 100. 

As it can be observed from Fig. \ref{curve1}, DeepCMC provides significant improvement in the quality of the reconstructed CSI at the BS with respect to CSINet and CRNet at all bit rate values. We remark here that, CSINet itself provides $3-6$dB improvement in $\mathrm{NMSE}$ compared to model-based CSI compression techniques in the literature exploiting sparsity of the channel gain matrix \cite{wen2018deep}. However, the gains from DeepCMC are even more drastic, achieving remarkably good reconstruction of the channel gain matrix with $\mathrm{NMSE}$ of $-13$~dB at a bit rate lower than $0.16$ bits per CSI entry. These results show that DeepCMC outperforms CSINet $4$ to $6$ dB in NMSE for the range of compression rates considered here. For example, for a target value of $\mathrm{NMSE}= -5$~dB, DeepCMC can provide more than 5 times reduction in the number of bits that must be fed back from the UE to the BS.

\subsection{Analog CSI feedback}

Analog CSI feedback follows a joint source-channel coding approach, and directly maps the downlink CSI to the uplink channel input in an unquantized and uncoded fashion. A CNN-based analog CSI feedback scheme, namely AnalogDeepCMC is proposed in \cite{mashhadi2019cnnbased}, to carry out the CSI compression and feedback tasks simultaneously, taking into account the feedback channel impairments. It uses a fully convolutional autoencoder model to efficiently map the downlink CSI at the UE to the uplink channel inputs, and to reconstruct them at the BS. The model is trained treating the uplink feedback channel as a non-trainable layer in the autoencoder. In this section, we provide performance comparisons between AnalogDeepCMC and the digital approach using DeepCMC for CSI compression, based on the quality of the reconstructed CSI at the BS when the same amount of uplink channel resources is devoted to CSI feedback. We will observe that the analog scheme improves the CSI reconstruction quality and consequently the achievable downlink rate without requiring the UL CSI at the UE for feedback transmission. 

Consider CSI feedback from a single-antenna user to a BS with $N_B$ antennas utilizing OFDM. Denote the uplink and downlink channel matrices by $\mathbf{H}_u \in \mathbb{C}  ^{K_u \times N_B}$ and $\mathbf{H}_d \in \mathbb{C}  ^{K_d \times N_B}$, respectively. Assume that the downlink CSI $\mathbf{H}_d$ available at the UE is fed back to the BS over $N_F$ uplink OFDM subcariers devoted to CSI feedback picked uniformly at random, with $\rho \triangleq \frac{N_F}{K_u}$ denoted as the \textit{feedback overhead}. The feedback channel over the $j$-th uplink subcarrier denoted by $\mathbf{{h}}_{F}^{j} \in \mathbb{C}  ^{N_B \times 1}, j = 1, \cdots, N_F$, is obtained from the corresponding row of $\mathbf{{H}_u}$, which specifies a SIMO channel with its output given by 
\begin{align}\label{FBCh1}
    \mathbf{y}_j= \mathbf{{h}}_{F}^{j} x_j + \mathbf{z}_j,
\end{align}
in which $\mathbf{y}_j \in \mathbb{C}  ^{N_B \times 1}$ is the received signal at the BS antennas, $x_j$ is the symbol fed back over the $j$-th subcarrier and $\mathbf{z}_j \in \mathbb{C}  ^{N_B \times 1}$ is the independent AWGN component. With $N_F$ uplink sub-carriers dedicated for CSI feedback, a maximum rate of $C_{FB}=\sum_{j=1}^{N_F}\log_2(1+SNR_{FB}\|\mathbf{{h}}_{F}^{j}\|^2)$ is available for CSI feedback, where $SNR_{FB}$ is the signal to noise ratio (SNR) in the uplink channel. However, note that $C_{FB}$ depends on the uplink channel state, which is not known by the UE. In a digital CSI feedback scheme, the UE will typically take a conservative approach and transmit at a rate that can be decoded with high probability. Using $C_{FB}$ as the feedback rate provides an upper bound on the performance of any digital CSI feedback scheme.

\begin{figure*}
\centering
\includegraphics[scale=0.5]{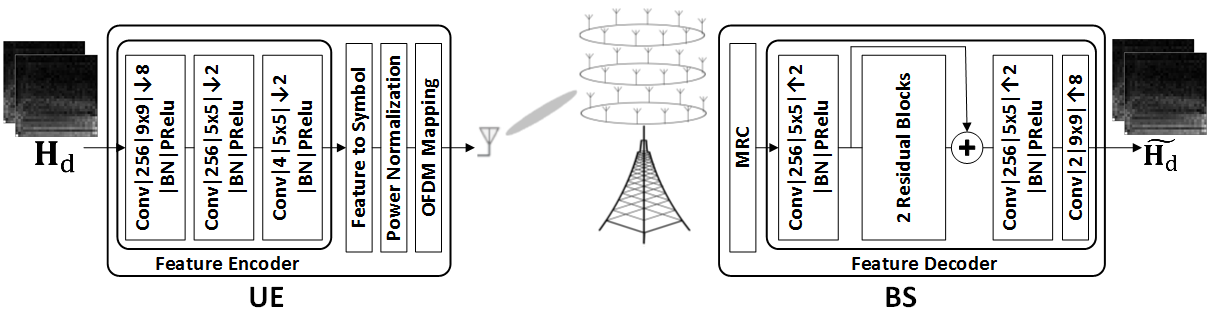}
\caption{AnalogDeepCMC for analog MIMO CSI feedback.}
\label{fig4}
\end{figure*}

Fig.~\ref{fig4} depicts the model architecture for Analog-DeepCMC~\cite{mashhadi2019cnnbased}. The UE applies a CNN-based feature encoder composed of three convolutional layers, which outputs real-valued features. Each pair of these real numbers are then grouped to form a complex-valued symbol, which are subsequently normalized to ensure the input power constraint over the feedback channel is met. These normalized symbols are then directly mapped to the corresponding subcarriers, and transmitted over the CSI feedback channel. The BS then performs MRC and feature decoding to reconstruct the original CSI matrix $\mathbf{H}_d$.  AnalogDeepCMC is trained including the feedback channel (noise and fading) as well as the MRC block as non-trainable layers in between the autoencoder structure. 

For comparison of the digital and analog feedback performance results, we will assume that the digital feedback can be provided at the instantaneous capacity of the feedback channel. 
As we have discussed above, this would require the UE to know the uplink channel state. However, during downlink channel training in an FDD MIMO scenario, the UE does not yet know the uplink CSI, and hence, will typically take a cautious channel coding and modulation approach which works at a rate significantly below $C_{FB}$. Moreover, we assume error-free transmission at the capacity despite a codelength of only $N_F$ symbols. Therefore, the corresponding NMSE result presented for the digital feedback scheme can be treated as a rather generous lower bound on the actual NMSE performance of any practical digital CSI feedback scheme.

\begin{figure}
\centering
\includegraphics[scale=.7]{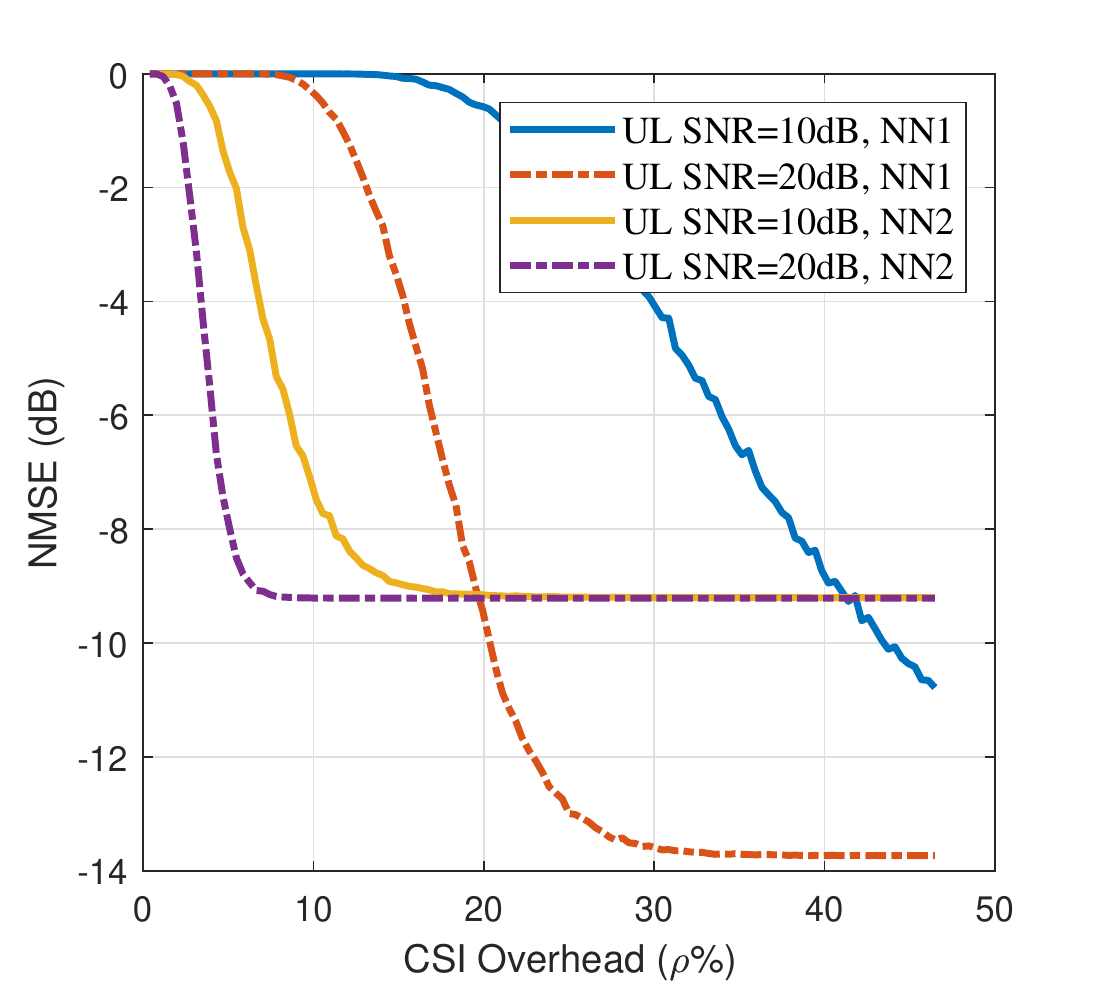}
\caption{Average NMSE vs. $\rho$ for digital CSI feedback using DeepCMC followed by capacity-achieving channel code.}
\label{curve2}
\end{figure}

Fig. \ref{curve2} depicts the average NMSE (dB) as a function of the CSI overhead $\rho$ for different values of the uplink channel SNR based on digital feedback using separate CSI compression with the DeepCMC algorithm followed by capacity-achieving channel coding. In this figure, DeepCMC is trained for two different $\lambda$ values resulting in NN1 and NN2. NN1 corresponds to a point with better reconstruction quality at a higher rate on the rate-distortion curve given in Fig. \ref{curve1}. The simulation scenario is the same as in Fig. \ref{curve1} with $K_d=256, N_B=32, N_U=1$. Note that DeepCMC is a variable-length lossy compression scheme; that is, for each CSI matrix, the UE obtains different number of bits at the output of the entropy coder. On the other hand, the capacity of the feedback channel is also random, depending on the states of the $N_F$ subcarriers dedicated to CSI feedback. Therefore, if the number of bits at the encoder's output exceeds $C_{FB}$, the feedback channel fails to deliver the CSI, called an \textit{outage} event, and the NMSE will equal $0$dB. This is why the NMSE curves all saturate at $0$dB for low $\rho$ values. If the CSI overhead decreases below a threshold, outages will occur with increasing frequency resulting in an increased NMSE. As $\rho$ increases beyond this threshold value, outage probability decreases with $\rho$. Beyond another higher threshold value, outage probability approaches zero, and the autoencoder reconstructs the CSI at the NMSE that it has been trained for (depending on the $\lambda$ value which controls the rate-distortion trade-off). This is the reason why the NMSE curves also saturate at high $\rho$ values. According to the figure, as the uplink SNR decreases, thresholds for both saturation regions increase. We would like to highlight that, for the setting considered here ($K_u=256$), $\rho=20\%$ would correspond to a channel code of length 51 symbols, in which case the code rates with reasonable reliability are significantly below the capacity \cite{polyanskiy2010channel}; that is, the NMSE values in this figure are quite generous for the digital scheme. 

As observed in Fig.~\ref{curve2}, for efficient digital CSI feedback, the UE requires the uplink CSI not only to decide on the appropriate channel coding rate, but also to use a NN trained with the proper $\lambda$ value to achieve the minimum possible NMSE. Networks trained for different reconstruction qualities result in different threshold behaviours. A network trained for better reconstruction quality results in an increased performance threshold but achieves a smaller NMSE for overhead values above the threshold. If uplink CSI is not available, which is the case during downlink training of FDD massive MIMO, the UE will typically need to take conservative source and channel coding approaches, which will result in considerable degradation of the CSI reconstruction quality with respect to those presented in Fig.~\ref{curve2}.  

\begin{figure}
\centering
\includegraphics[scale=.7]{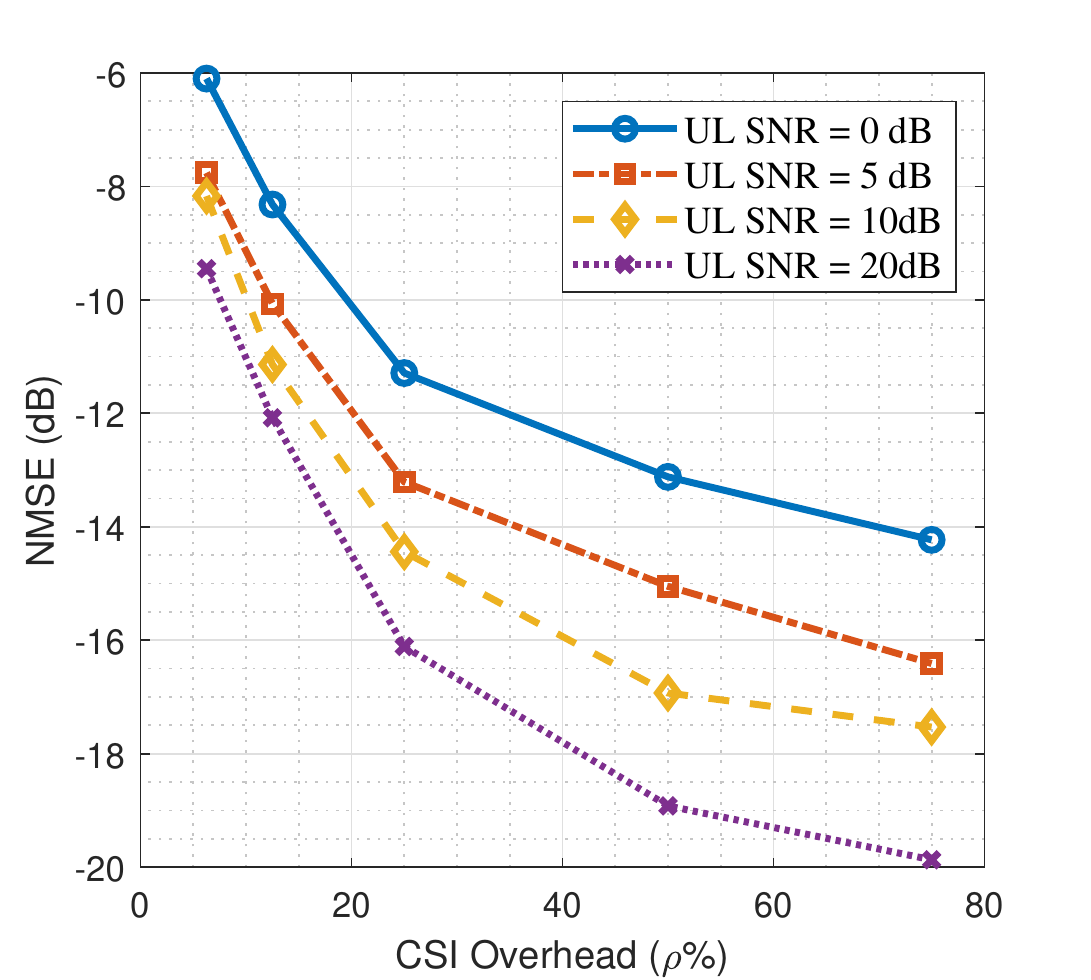}
\caption{Average NMSE vs. $\rho$ for the AnalogDeepCMC architecture.}
\label{curve3}
\end{figure}

Fig.~\ref{curve3} depicts the average CSI reconstruction NMSE (dB) as a function of the CSI overhead $\rho$ for different values of the uplink SNR using AnalogDeepCMC. The curves for different SNR values in Fig.~\ref{curve3} correspond to NN models trained for the corresponding uplink SNR. As observed in Fig. \ref{curve3}, there is no threshold behaviour in the analog CSI scheme and the NMSE curves exhibit graceful performance degradation with decreasing SNR in the uplink channel. This is unlike the digital feedback approach, which may result in severely degraded CSI quality (NMSE= $0$dB) due to outages if the uplink SNR decreases below a threshold. Hence, unlike the digital CSI scheme, AnalogDeepCMC does not require uplink CSI to send the downlink CSI back to the BS. The analog CSI scheme is much more favourable not only due to avoiding the performance thresholds and eliminating the need for explicit uplink CSI, but also for avoiding the channel coding and modulation delays. 

\section{CSI Training with Side Information}\label{sec4}

In the previous sections, we have focused on exploiting the joint distribution of CSI matrices to reduce the overhead for CSI estimation and feedback using NNs for lossy CSI compression. In this section, we will explore how we can exploit joint distribution across antennas and subcarriers, or across time and space to further reduce, or even completely remove the amount of required CSI feedback. This is based on the idea of using the available CSI information at the BS at a certain point in time, space or frequency or a subset of antennas as correlated side information to improve the compression efficiency with NNs. This can be considered as implementing Wyner-Ziv lossy compression \cite{WynerZiv} in the presence of correlated side information at the receiver. 

As an example, consider a FDD massive MIMO scenario, where channel reciprocity does not hold, and separate downlink and uplink training would normally be necessary. Although the uplink and downlink channels are not fully reciprocal, the uplink and downlink signals traverse the same geometrical paths with different frequencies, which imposes some correlation between them. The authors in \cite{DLCSI4} use the uplink CSI (which is already available at the BS by uplink training) as a side information to further improve downlink CSI reconstruction performance by utilizing the correlations between downlink and uplink channels. In \cite{FEDDEL}, the authors use delayed CSI (which has been delayed due to the limited  communication rate in the feedback channel) as the correlated side information. As another example, the BS can exploit joint distribution of the CSI for nearby UEs to estimate, compress and feedback the CSI jointly at a reduced overhead. Considering downlink training of a FDD massive MIMO system, the authors in \cite{yang2019deepJ, Cooperative} use a NN to learn and exploit joint distribution of the CSI for nearby UEs and the correlation among their channels to reduce the CSI feedback overhead similarly to a distributed lossy compression scheme.  

If the side information proves to be sufficient for predicting the required CSI with an acceptable distortion using a NN, then the NN can characterize a mapping function to predict the required CSI from the available side information with zero overhead. Such mappings can significantly reduce the CSI acquisition overhead and have been considered in \cite{Mapping} for different mapping scenarios in frequency and space. As an example, the authors in \cite{Mapping, ULDLLet} train a fully connected NN to predict downlink from the uplink CSI, and hence totally eliminate the downlink training and feedback overhead.

\section{Conclusion and future research directions}\label{sec5}

Massive MIMO systems are considered as the key technology to enable the excessive throughput requirements in 5G and future generation wireless networks due to their ability to serve many users simultaneously with high spectral and energy efficiency. However, due to the drastic increase in the number of antennas, CSI acquisition and feedback become challenging, requiring excessive time, frequency and computational resources potentially crippling benefits of massive MIMO systems. Many previous works have taken model-driven approaches assuming sparse or low-rank models on the CSI matrix to reduce the overhead. However, these techniques cannot exploit statistical structures that go beyond sparsity. This encouraged data-driven approaches based on training NN architectures over large datasets of CSI matrices, generated using accurate channel models or even from channel measurements, to capture these structures and use them to reduce CSI acquisition and feedback overhead. DL-based approaches have shown significant improvements in comparison with traditional methods for CSI estimation, compression and feedback. Yet, there is still much room for future research.

Many NN architectures proposed so far use fully-connected layers for different estimation/compression tasks. While  fully-connected NNs have the potential to learn and exploit complex joint distributions across all the antennas and subcarriers, they do not easily scale with MIMO dimensions, and need separate training for different number of antennas, subcarriers, etc. Hence, more insights on the correlation structures of the CSI in practical massive MIMO systems need to be exploited to design more efficient and less complex NNs for the CSI acquisition tasks. On the other hand, many of the existing works focus on a single task (e.g., channel estimation) and propose a NN architecture to achieve optimized performance for that specific task. However, the complete CSI training process consists of pilot transmission, CSI estimation, and feedback, and these tasks interact and effect each other (e.g., a less accurate channel estimate may be good enough if the subsequent CSI compression block would introduce significant distortion due to the limited capacity of the feedback channel). On the other hand, NNs have the capability to model and optimize processes in an end-to-end manner. A NN architecture trained end-to-end for the CSI acquisition task is not yet available. We believe such an end-to-end optimized architecture will not only benefit from the interactions between different tasks, but also can reduce the overall complexity of the NN by avoiding repeated layers that would be required in a task-by-task design process. We also note that, most of the existing results consider a single-user or a single-cell massive MIMO system, and extending these results to more general multi-cell multi-user massive MIMO scenarios is another potential direction for future research. Finally, practical implementation of these NN-based techniques in real environments on real devices has not yet been studied, but may be critical not only to evaluate the performance of the NNs trained on data generated by various channel models in real channel conditions, but also to understand the impact of limited computational resources available at the UEs.

\bibliographystyle{IEEEtran}
\bibliography{refs}
\end{document}